\documentclass[aps,pra,10pt]{revtex4-2}
\usepackage{graphicx}
\usepackage{dcolumn}
\usepackage{bm}
\usepackage{lmodern}
\usepackage{mathtools,amsmath,amssymb}
\usepackage{dsfont}
\usepackage{mathrsfs}
\usepackage{amsthm,amsfonts,mathtools}
\usepackage{mathbbol}
\usepackage{hyperref}
\usepackage{xcolor}
\usepackage{subfigure}

\newcommand{\rrho}{\hat{\boldsymbol{\rho}}}
\usepackage[normalem]{ulem}

\newcommand{\Tr}{\mathrm{Tr}}

\newcommand{\reals}{\mathds{R}}
\newcommand{\complex}{\mathds{C}}
\newcommand{\Rh}{\hat{\boldsymbol{\rho}}}
\theoremstyle{plain}

\theoremstyle{definition}

\begin{document}
\title{Overcoming sloppiness for enhanced metrology in continuous-variable quantum statistical models}	

\author{Massimo Frigerio}
\email{Electronic address: massimo.frigerio@lkb.upmc.fr}
\affiliation{Laboratoire Kastler Brossel, Sorbonne Université, 4 place Jussieu, 75005 Paris, France}

\author{Matteo G. A. Paris}
\email{Electronic address: matteo.paris@fisica.unimi.it}
\affiliation{Dipartimento di Fisica,
Universit\`a  di Milano, I-20133 Milano, Italy}

\begin{abstract}
Multi-parameter statistical models may depend only on some functions of the parameters that are fewer than the number of initial parameters themselves. Such \emph{sloppy} statistical models are characterized by a degenerate Fisher Information matrix, indicating that it is impossible to simultaneously estimate all the parameters. In a quantum setting, once an encoding is fixed, the same can happen for the Quantum Fisher Information matrix computed from a sloppy quantum statistical model. In addition to sloppiness, however, further issues of quantum incompatibility can arise. We take a fully Gaussian case-study to investigate the topic, showing that by appropriately scrambling the quantum states in between the encoding of two phase-shift parameters a Mach-Zehnder interferometer, not only sloppiness can be lifted, but also the quantum incompatibility can be put identically to zero, maintaining an enhanced scaling of precision and the covariance of the model with respect to exact values of the parameters. 
\end{abstract}
\maketitle

\section{Introduction, motivations and layout}
The rapidly developing field of quantum metrology and quantum sensing \cite{PARIS2009,dowling15,taylor16,montenegro24}, seeking quantum enhancements in parameter estimation, draws inspiration from the fragility 
of quantum states to improve the results of classical metrology by suppressing 
errors below the shot noise fluctuations, typically proportional to the square 
root of the number of particles involved, thus achieving greater precision than what would be allowed by classical probes and classical measurements. 

The minimal uncertainty achievable for quantum systems, quantified by the 
quantum Cramér-Rao bound (CRB), is in general achievable only for single-parameter 
models and adequate choice of the estimators, whereas extending it to multi-parameter 
models poses significant challenges \cite{Genoni13,szczykulska16,wang20,Albarelli2020}, e.g. 
if the optimal measurements for the 
estimation of one parameter do not commute with those for another, thus 
showcasing an intrinsically quantum hindrance. Fixed measurements may induce 
parameter correlations, necessitating trade-offs in uncertainty distributions 
among parameters, complicating multiparameter optimization. For these reasons, 
most neat results in quantum metrology are applicable only to single-parameter
estimation. Yet, the multifaceted nature of certain systems, like biological 
samples or nonhomogeneous crystals, requires a multi-parameter approach \cite{transtrum15}. 

In parallel to the quantum incompatibility issues, there could also be sources 
of classical indeterminacy in the independent estimation of different parameters. 
This happens when the encoding of the parameter on the statistical model is such 
that the resulting probability distributions only depend upon certain functions 
of the parameters, that are fewer in number with respect to the initial parameters 
themselves \cite{waterfall06,gutenkunst07}. In this case, through a 
(possibly nonlinear) invertible reparametrization, 
it will be possible to rewrite the model in terms of a smaller number of parameters. 
Since the Fisher Information matrix transforms by congruence with the Jacobian of 
the reparametrization functions, its eigenvalues are invariant. We may thus \emph{define} 
a sloppy statistical model as one whose Fisher Information matrix has some zero eigenvalues. 
In a more empirically relevant form, an upper bound would be placed to the eigenvalue
for the corresponding parameter to be considered relevant to the model, and call 
\emph{sloppy} any model whose FI matrix has at least one eigenvalue lower than 
this threshold. The same definitions straightforwardly apply also to quantum statistical 
models and Quantum Fisher Information matrix, since the latter is just a special 
case of a Fisher Information matrix for the optimal measurement. 

In this article, we face the problem of estimating two consecutive phase shifts acting 
along one arm of a Mach-Zender interferometer and focus the analysis to Gaussian 
states and operations. The situation considered is rather simple, yet it may provide good modeling of optical media with a spatially variable refractive index, which may results from the structural characteristic of the material (e.g., layered media) or from a possible temperature dependence \cite{Meyzonnette2019}.

{After a short review in Sect. \ref{sec:I} of the main concepts and theoretical tools of multi-parameter quantum metrology, in Sect. \ref{sec:II} we introduce the model that will be studied and we compute the diagonal elements of the quantum Fisher information matrix for the two phase parameters: by feeding the interferometer with two identical single-mode squeezed vacuum states with a relative displacement and adding another squeezer in between the two phases, we seek the optimal setting of the interferometer to reach the highest Quantum Fisher Information for the two phase parameters. We then tackle the issues of sloppiness and quantum incompatibility of such a model in Sect. \ref{sec:III}, arriving at results in closed form that shine light into the largely unexplored topic of sloppy quantum statistical models. Our goal is to lift the sloppiness but keeping the model quantum-compatible, so as to ensure that the symmetric logarithmic derivative Quantum Cramér-Rao bound (SLD-QCRB) can still be saturated, at least in principle. Our results prove that concentrating quantum resources on the optical element to be estimated is in general the best choice to enhance precision, while the use of entanglement in the probe state is instead better at estimating the two parameters separately.}

\section{Multi-parameter quantum metrology}
\label{sec:I}
The general theoretical framework to deal with estimation problems of continuously 
varying parameters is that of statistical models. In the classical case, these are 
defined as families of probability density function $p_{\vec{\lambda}}: X \rightarrow [0,1]$ with respect to some continuous or discrete variable $x \in X$, continuously labelled by real parameters $\vec{\lambda}$ belonging to an open subset $\Lambda \subseteq \reals^n$. Prescribing this dependence upon the parameters amounts to fixing the statistical model. The goal of estimation theory is to estimate the true value of the parameters, and therefore the true probability density function inside the family, by having a finite set of $M$ data points sampled from it and using some estimator $\Theta^{(M)}: X^{M} \to \Lambda$, i.e. a function of the dataset that gives the string of estimated parameters as the output. If this estimator is unbiased, meaning that its expectation value coincides with the true value of the parameters vector:
\[  \mathrm{E}_{\vec{\lambda}} \left[  \Theta^{(M)} \right] = \vec{\lambda}  \]
then the covariance matrix of the parameters estimated with it:
\[ \Sigma_{jk} ( \Theta^{(M)} ) \ = \ \mathrm{E}_{\vec{\lambda}} \left[ \left( ( \Theta^{(M)})_{j} - \lambda_{j} \right)   \left( ( \Theta^{(M)})_{k} - \lambda_{k} \right)        \right] \]

is bounded by the \emph{Cramér-Rao bound}:
\begin{equation}
    \label{eq:CramerRao}
    \Sigma \left( \Theta^{(M)} \right) \ \ \geq \ \ \ \frac{1}{M} \ \mathscr{F}^{-1} [  \vec{\lambda}^{*}]
\end{equation}
where the functional $\mathscr{F}$, known as the \emph{Fisher Information Matrix}, associates a positive-definite invertible matrix to each probability distribution in the statistical model, computed as:

\begin{equation}
    \mathscr{F}_{jk} [ \vec{\lambda} ] \ = \  \int_{X} \ p_{\vec{\lambda}} \  \partial_{{j}}  \left[ \log p_{\vec{\lambda}} \right]      \ \partial_{{k}} \left[ \log p_{\vec{\lambda}} \right] \  dx
\end{equation}
using the short-hand notations $p_{\vec{\lambda}} \equiv p_{\vec{\lambda}} (x)$ and $ \partial_{{j}} \equiv  \partial_{\lambda_{j}} $. 
An asymptotically efficient estimator, i.e. one that saturates the Cramér-Rao inequality, at least asymptotically, i.e., for $M \to +\infty$, always exists and it is provided by the maximum-likelihood estimator and by the Bayesian estimator. 

In quantum mechanics, however, this family of probability distributions can only be obtained after specifying a measurement setup. To be able to discuss quantum estimation without picking a specific measurement from the start, it is useful to consider \emph{quantum statistical models}, which are maps from the open set $\Lambda \subseteq \reals^n$ of real parameters to quantum states on some Hilbert space $\mathcal{H}$, $\vec{\lambda} \mapsto \rrho_{\vec{\lambda}} \in \mathcal{T}( \mathcal{H} ) $, where $\mathcal{T}$ denotes the set of trace-class, bounded, positive-semidefinite linear operators. As long as the map is continuous, $\Lambda$ is open and the operators in the model are all true quantum states (i.e. they do belong to $\mathcal{T}$), one can always define the \emph{Quantum Fisher Information} (QFI) matrix as:
\begin{equation}
\label{eq:SLDtoQFI}
  \mathcal{Q}_{jk} ( \vec{\lambda} ) \ = \ \Tr \left[ \Rh_{\vec{\lambda}} \,\dfrac{\hat{\mathrm{L}}^{S}_{j} \hat{\mathrm{L}}^{S}_{k} + \hat{\mathrm{L}}^{S}_{k} \hat{\mathrm{L}}^{S}_{j}}{2}   \right]
\end{equation}
where the Symmetric Logarithmic Derivative (SLD) operators $\hat{\mathrm{L}}^{S}_j$ are implicitly given by the solution of the following Lyapunov Equation:
\begin{equation}
    \label{eq:SLDmultiparam}
    \partial_{\lambda_{j}} \Rh_{\vec{\lambda}} \ = \ \dfrac{ \hat{\mathrm{L}}^{S}_{j} \Rh_{\vec{\lambda}} + \Rh_{\vec{\lambda}} \hat{\mathrm{L}}^{S}_{j}}{2}
\end{equation}
The Quantum Fisher Information always results in a valid Cramér-Rao bound, i.e., 
\begin{equation}
    \label{eq:qCramerRao}
    \Sigma \left( \Theta^{(M)} \right) \ \ \geq \frac{1}{M} \ \mathcal{Q}^{-1} [  \vec{\lambda}^{*}]\,.
\end{equation}
However, this inequality is in general not saturable for multi-parameter models, since there could be no single measurement providing a probability distribution from which all parameters can be simultaneously estimated. This is in contrast with the simpler case of one-parameter quantum statistical models, for which the eigenprojectors of the single SLD operator define the best measurement, i.e. the one providing the highest FI value, and the bound can in turn always be saturated. Other QFI matrices can be defined by using other definitions of the Logarithmic Derivative, but none of them provides a tight bound that can be saturated for the multi-parameter case. 

All those matrix inequalities give raise to scalar bounds by introducing a {\em weight matrix}, i.e., a $n\times n$ semipositive matrix $\bf W$, and taking the trace of Eqs. (\ref{eq:CramerRao}) and (\ref{eq:qCramerRao}). We have \cite{Albarelli2020}
$$\Tr[{\bf W \, \Sigma}] \ge  C_{\mathcal{F}}({\bf W}) 
\qquad \Tr[{\bf W \, \Sigma}]\ge C_{\mathcal{Q}}({\bf W}),$$ where 
$$C_{\mathcal{F}} ({\bf W})= M^{-1} \, \Tr[{\bf W} \, \mathcal{F}^{-1}] \qquad
C_{\mathcal{Q}} ({\bf W})= M^{-1} \,\Tr[{\bf W}\, \mathcal{Q}^{-1}].$$ 

In the single-parameter scenario, the bound (\ref{eq:qCramerRao}) can be achieved by a projective measurement over the SLD eigenstates, while in the multi-parameter setting,
it is not attainable in general, as the SLDs associated with the different parameters may not commute with one another. In this case, the parameters are incompatible, and there is no joint measurement that allows one to estimate all the parameters with the ultimate precision. 
In this case, two other relevant bounds may be introduced. The first is the so-called {\em most informative bound}, $C_{\rm MI}({\bf W})= M^{-1} \min_{\mathbf{\Pi}} \{\Tr[{\bf W} \, \mathcal{F}^{-1}]\}$, minimized over all possible measurements $\mathbf{\Pi}$. The quantity
$C_{\rm MI}({\bf W})$ does not, in general, coincide,  with $C_{\mathcal{Q}}({\bf W})$ in the multiparameter case. The so-called Holevo Cramér-Rao (HCR) bound  $C_{\rm H}({\bf W})$\cite{Holevo1977} corresponds to the most informative bound of the asymptotic statistical model, i.e. the minimum FI bound 
achieved by a collective measurements performed on infinitely many copies of the statistical model \cite{Albarelli2020, Razavian20}. In turn, we have $\Tr[{\bf W \, \Sigma}] \ge  C_{\mathcal{F}}({\bf W})  \ge  C_{\rm MI}({\bf W}) \ge  C_{\rm H}({\bf W}) \ge C_{\mathcal{Q}}({\bf W})$. Therefore, the tightest bound, at least asymptotically and with collective strategies on multiple copies of the unknown state, is provided by the Holevo bound. 

The Holevo bound $C_{\rm H}({\bf W})$ is usually difficult to evaluate compared to $C_{\mathcal{Q}}({\bf W})$ and therefore the following relation represents a useful tool in characterizing the a multiparameter estimation model
\begin{eqnarray}\label{eq:HolevoHier}
C_{\mathcal{Q}}({\bf W}) \le C_{\rm H}({\bf W}) \le (1+{\cal R}) C_{\mathcal{Q}}({\bf W}) \, ,
\end{eqnarray}
where the {\em quantumness} parameter ${\cal R}$ 
is given by \cite{carollo20,Carollo2019,candeloro21} 
\begin{eqnarray}\label{eq:R}
{\cal R} = \| i \, \mathcal{Q}^{-1} \mathcal{U} \|_\infty\, .
\end{eqnarray}
In the above equation, $\|{\bf A}\|_\infty$ denotes the largest eigenvalue 
of the matrix $\bf A$, and
$\mathcal{U} (\bf \lambda)$ is the asymptotic incompatibility matrix, also referred to as {\em Uhlmann curvature}, with matrix elements \cite{carollo20}:
\begin{eqnarray}\label{eq:Uhl}
{\mathcal{U}}_{\mu \nu}= -\frac{i}{2} \Tr \left\{\rho_{\bf\lambda} [L_\mu, L_\nu] \right\} \, ,
\end{eqnarray}
where $[A,B]=AB-BA$ is the commutator of $A$ and $B$
{Equation~(\ref{eq:HolevoHier}) implies that QFI bound may be saturated iff ${\mathcal{U}} (\bf\lambda)=0$, which is usually referred to as the {\em weak compatibility condition}.}
The quantumness parameter ${\cal R}$ is bounded $0\le {\cal R} \le 1$ and vanishes ${\cal R}=0$ iff $\mathcal{U} (\bf\lambda)=0$. Therefore, it provides a measure of asymptotic incompatibility between the parameters. For $n=2$ we may write
\begin{eqnarray}\label{eq:R1}
{\cal R} = \sqrt{\frac{\det \mathcal{U}}{\det \mathcal{Q}}} \qquad \mbox{for $n=2$ parameters.}
\end{eqnarray}

For pure Gaussian models (see Appendix \ref{appx:Gauss} for conventions), as those we are going to deal in the following, it is possible to derive a simple expression for the Quantum Fisher Information \cite{Pinel12,monras13,jiang14,Safranek15,sparaciari15,marian16}:
\begin{equation}
\label{eq:QFIGauss}
\begin{aligned}
    \mathcal{Q}_{jk} \ = \ & \frac{1}{4} \Tr \left[ \left( \sigma^{-1}  \partial_{j} \sigma \right) \left( \sigma^{-1}  \partial_{k} \sigma     \right)   \right] + 2 ( \partial_{j} \vec{d}^{T}  ) \cdot \sigma^{-1} \cdot  ( \partial_{k} \vec{d} )
\end{aligned}
\end{equation}
Here $\sigma \equiv \sigma_{\vec{\lambda}}$ is the covariance matrix and $\vec{d} \equiv \vec{d}_{\vec{\lambda}}$ is the mean-field vector of the pure Gaussian state $\lvert \psi_{G} (\vec{\lambda}) \rangle$ defining the model and $\partial_{j} \equiv \partial_{\lambda_{j}}$. Remarkably, this expression is the same as the classical Fisher Information for a classical statistical model consisting of Gaussian probability density functions with CM $\sigma_{\vec{\lambda}}$ and expectation values $\vec{d}_{\vec{\lambda}}$. 
Besides Eq.(\ref{eq:QFIGauss}) for the QFI, it is possible to derive an analogous expression for the Uhlmann curvature of a pure Gaussian quantum model \cite{Safranek15}:
\begin{equation}
\label{eq:UGauss}
\begin{aligned}
    \mathcal{U}_{jk} ( {\vec{\lambda}}) \ =   & \ \frac{1}{4} \Tr \left[ \Omega \sigma \left( \Omega \partial_{i} \sigma \Omega \partial_{j} \sigma \ - \   \Omega \partial_{j} \sigma \Omega \partial_{i} \sigma \right) \right] +  4 \partial_{i} \vec{d}^{ \ T} \sigma^{-1} \Omega \sigma^{-1} \partial_{j} \vec{d}
\end{aligned}
\end{equation}

\section{Sloppy models in Mach-Zehnder interferometry}
\label{sec:II}
A multi-parameter statistical model, be it classical or quantum, can sometimes be not invertible in an open neighbourhood around a point in parameters space, meaning that different parameters values are assigned locally to the same state and it is impossible to accurately recover them from the statistics. In these cases, we say that the (classical or quantum) statistical model is \emph{sloppy}. The degeneracy of the (quantum) Fisher Information matrix signals that there is some reparametrization, not necessarily a linear one, such that, in terms of the new parameters, one or more of them have an almost vanishing value of the (quantum) FI and the statistical model is effectively only dependent upon the other parameters. Of course, locally in the parameters space any reparametrization is linear on the FI matrix because it acts by congruence on it, but the Jacobian can depend upon the values of the parameters.

In this framework, it should be intuitively clear that the sloppiness of a statistical 
model is an issue about the \emph{encoding} of the parameters into the set of states. 
For example, imagine that we want to estimate the parameters of a CV unitary operator 
acting on a single-mode field by first rotating its quadratures by an angle $\theta$ 
and then displacing the mean field by $\alpha \in \complex$. If we decide to encode 
$(\theta, \alpha)$ on a thermal state, the dependence upon $\theta$ will be completely 
lost: the obvious option would be to change the probe state into another which does not commute with $e^{i \theta \hat{a}^\dagger \hat{a}}$. Clearly, if the agents determining 
the two parameters can be addressed independently, we are not actually dealing with a 
true two-parameters model and we could just estimate them individually with
 single-parameter quantum metrology: in this section, therefore, we assume to 
 ave \emph{limited control} about the single agents.

\begin{figure}[htb!]
    \centering
    \includegraphics[width=0.96\columnwidth]{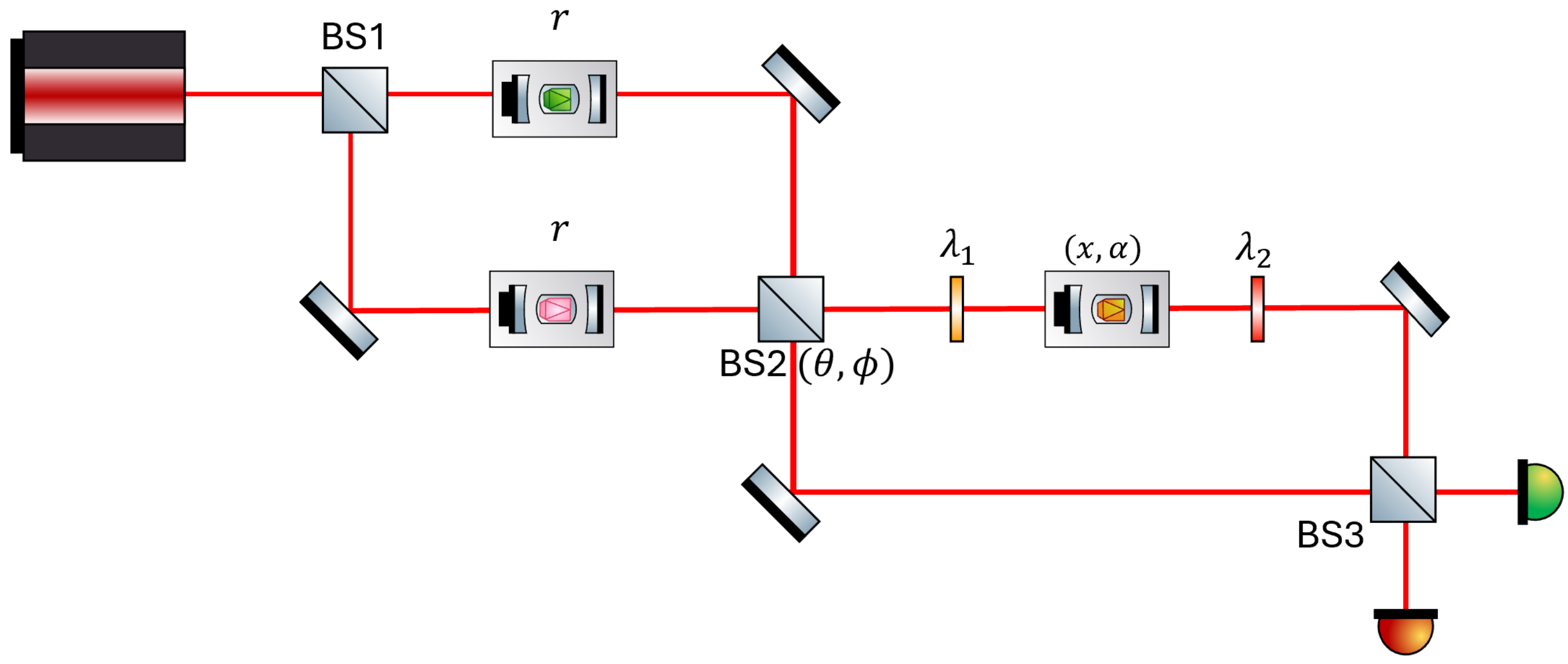}
    \caption{Two single-mode squeezed vacuum states, each with the same squeezing parameter $r$, are injected into a Mach-Zehnder interferometer. The beam splitter has a transmissivity of $\cos^{2} \phi$, and insert a relative phase $\theta$ between the input modes. The parameters $\lambda_{1}$ and $\lambda_{2}$, which are to be estimated, are encoded by two consecutive phase shifters on the upper arm of the interferometer. An additional single-mode squeezer, with squeezing parameter $x$ and squeezing angle $\alpha$, is placed between the two phase shifters to decouple them and distribute information over a larger portion 
of the Hilbert space.}
     \label{fig:MachZendersloppy}
\end{figure}

In order to address the issue of sloppiness with some generality while still obtaining analytic results, we examine the fully Gaussian, two-mode sloppy model illustrated in Fig. \ref{fig:MachZendersloppy}. Our goal is to investigate the relationship between sloppiness and quantum incompatibility. We consider a Mach-Zehnder interferometer fed with two independent and equally squeezed vacuum states, whose covariance matrix (CM) is given by:

\begin{equation}
 \sigma_{0}^{(1)} \ \ = \ \ \frac{1}{2} \left(\begin{array}{cccc}  e^{2r} & 0 & 0 & 0 \\
                                            0 & {e^{-2r}} & 0 & 0 \\
                                            0 & 0 & e^{2r} & 0 \\
                                            0 & 0 & 0 & e^{-2r} \end{array} \right)  
\end{equation}
and zero mean-field vector. This is obtained by splitting the laser source on a beam splitter (BS1) and let the two resulting beams act as pumps for two nonlinear crystals. The resulting two mode state, $\Rh^{(0)}_{1,2} =  \Rh^{(0)}_{1;r} \otimes \Rh^{(0)}_{2;r}$, is therefore the tensor product of two single-mode squeezed vacuum states with the same squeezing parameter $r$. The first beam splitter of the interferometer (BS2) has mixing angle $\phi$, such that the transmissivity is $\cos^{2} \phi$, and applies a relative phase $\theta$
. 
A relative phase in the squeezing angle of one of the input states can be reabsorbed in $\theta$. The parameters to be estimated are the phase shifts $\lambda_{1}$ and $\lambda_{2}$ imprinted on the mode occupying the upper horizontal arm of the interferometer by two consecutive phase shifters. The second beam splitter, the mirrors and everything afterwards are irrelevant as long as the Quantum Fisher Information is concerned, as they amount to a parameter-independent unitary change of the quantum statistical model, which is reabsorbed when considering all possible POVMs. 

When the squeezer in between the two phase shifters is absent ($x=0$), the model is sloppy since the two parameters contribute to an overall phase shift of $\lambda_{1} + \lambda_{2}$ of one mode with respect to the other, while the difference $\lambda_{1} - \lambda_{2}$ cannot be retrieved by any measurement on the two-mode output state. {A realistic scenario described by such a model that we can have in mind to ponder on it could be a continuous medium of unknown and continuously varying refractive index that imparts a total phase shift and we would like to estimate how different sections of it contribute to the overall shift.} 

If we limit our analysis to Gaussian states, the only nontrivial decoupler that could be inserted between the two phase shifters is a single-mode squeezer, parametrized by a squeezing factor of $e^{-2x}$ and a squeezing angle $\alpha$, as in Fig.\ref{fig:MachZendersloppy}, and possibly an additional displacement. An intermediate displacement would be in principle sufficient to partially uncouple the two parameters, but the discrimination of $\lambda_{1}$ and $\lambda_{2}$ would then be limited to the first-moments vector of the final Gaussian state, thus resulting in uninteresting conclusions. 

Overall, we can write the general output state as:
\begin{align}
    \label{eq:QSMsloppy}
     \hat{\mathbf{U}}_{\phi, \theta, x , \alpha} ( \lambda_{1}, \lambda_{2} ) &  :=  \ \hat{\mathbf{U}}_{R} ( \lambda_{2} )  \ \hat{\mathbf{S}}^{(1)}_{x, \alpha} \ \hat{\mathbf{U}}_{R} ( \lambda_{1} ) \hat{\mathbf{U}}_{BS}( \theta, \phi) \\
   \Rh (\lambda_{1}, \lambda_{2} ; r,  \phi, \theta, x, \alpha, q, \beta ) 
   & =    \hat{\mathbf{U}}_{\phi, \theta, x , \alpha} \left[ \Rh^{(0)}_{1;r,q, \beta} \otimes \Rh^{(0)}_{2;r,0,0}         \right] \left( \hat{\mathbf{U}}_{\phi, \theta, x , \alpha} \right)^{\dagger}
\end{align}
where $\hat{\mathbf{U}}_{R}(\lambda) = \exp \left[- i \lambda \hat{a}^\dagger \hat{a}\right]$ and
\begin{equation}
\hat{\mathbf{S}}^{(1)}_{x, \alpha} := \exp \left[ \frac{x}{2} \left( e^{i \alpha} \hat{a}^{\dagger 2} - e^{- i \alpha} \hat{a}^2 \right) \right] \ , \ \ \hat{\mathbf{U}}_{BS}( \theta, \phi) := \exp \left[ i \theta \left( e^{i \phi} \hat{a}^\dagger \hat{b} +e^{-i \phi} \hat{b}^\dagger \hat{a}  \right) \right]
\end{equation} 
and we omitted the explicit dependence of $\hat{\mathbf{U}}_{\phi, \theta, x , \alpha}$ on $(\lambda_{1}, \lambda_{2})$ for brevity. 
Notice that  the initial sloppy model is fully covariant with respect to $\lambda_{1}$ and $\lambda_{2}$, 
This means that the (degenerate) Quantum Fisher Information (QFI) matrix does not depend on the actual values of these parameters. In contrast, the model with the scrambler in Eq.(\ref{eq:QSMsloppy}) is only covariant with respect to $\lambda_{2}$. Additionally, it is rather clear that $\lambda_{1}$ and $\alpha$ only enter in the model through the combination $\gamma = \alpha + 2 \lambda_{1}$, meaning that the phase $\alpha$ can be exploited to improve the local estimation if the QFI around the true value of $\lambda_{1}$ is not optimal. 

To compute the SLD-QFI matrix of the QSM with decoupler, we resort to 
the formula in Eq.(\ref{eq:QFIGauss}), obtaining the following entries:

\begin{align}
   \mathcal{Q}_{11} (r, \phi, \theta, q, \beta ) & =  2 \cosh^2 2r +  2 q^2 \Big\{ \cosh 2 r  +  \\ \notag & \left[ \cos 2 \beta \cos 2 \phi \ +  \cos (2 \beta + \theta) \sin \theta  \sin(2 \phi)^2\right] \sinh
      2 r \Big\}
\\
        \mathcal{Q}_{22}(r,\theta, \phi, x, \gamma, q) & =   2 \Big( \cosh 2 r  
        \cosh 2 x \ + \ \sinh 2 r \ \left[ \cos \theta \cos (\gamma  + \theta)  
        \right. \\ & \notag \left. +  \cos 2 \phi \sin \theta \sin ( \gamma + \theta) \right] \sinh 2x \Big)^2  +  2 q^2 f_{22}(r,\beta, \theta,\phi,x,\gamma) 
\\
    \mathcal{Q}_{12}( r, \theta, \phi, x,  \gamma, q)  & =  2 \cosh^2 2r  
   + 2 \left[ 2 - \sin^2 (2 \phi ) \sin^2 \theta \right] \sinh^2 (2 r)\sinh^2 x 
     +  \\ & \notag \Big[ \cos\theta \cos(\gamma + \theta) \ + 
    \cos(2\phi) \sin \theta \sin(\gamma + \theta) \Big] \sinh 4r \sinh 2x  \\ \notag &  + 
    2 q^2 f_{12}(r,\beta,\theta,\phi,x,\gamma)
\end{align}
In the equations for $\mathcal{Q}_{22}$ and $\mathcal{Q}_{12}$ we left implicit the term dependent on $q$ since its expression is rather involved and uninformative, but it is reported in its entirety in Appendix \ref{appx:A}. We also reparametrized with $\gamma = \alpha + 2 \lambda_{1}$, since this is the only combination of these parameters appearing in the expressions. The $q$-independent term of $\mathcal{Q}_{11}$ is just $2 \cosh^{2} 2r$, while the $q$-dependent term can be maximized by putting $\beta = \theta = 0$ and $\phi ={ n \pi}$, or $\theta = \frac{\pi}{2}$, $\beta = - \frac{\pi}{4}$ and $\phi = \frac{(2n+1) \pi}{4}$, where $n$ is a positive integer. Its value at the maximum is:
\begin{equation}
    \mathcal{Q}_{11}^{max} \ = \ 2 \cosh^2 (2 r) + 2 e^{2r} q^2
\end{equation}
For constant input energy (i.e. constant $r$) we can now maximize $\mathcal{Q}_{22}$. The absolute maximum is achieved for $\phi = \theta = \gamma = 0$, dubbed the \emph{maximum configuration} from now on, and yields:
\begin{equation}
\label{eq:Q22max}
    \mathcal{Q}_{22}^{max} \ = \  2 \cosh^2 \left[ 2 ( r + x) \right]
\end{equation}
    and it is compatible with the maximum of $\mathcal{Q}_{11}$ even for $q \neq 0$, as can be seen by studying the behavior of the function $f_{22}$ provided in Appendix \ref{appx:A}. However, this result is of little relevance for at least two reasons: we had to put $\gamma= 0$, which amounts to fine-tuning $\alpha$ according to the true value $\lambda_{1}$ of the first unknown parameter, and we also end up with a solution corresponding to removing the initial beam-splitter from our scheme ($\phi = 0$). This fact has a clear interpretation: if we want to estimate $\lambda_{1}$ and $\lambda_{2}$ with the best possible accuracy \emph{overall} and \emph{disregarding their compatibility} we have to transmit all the incoming squeezing onto the upper arm of the interferometer where the phase shifters are placed. To be more explicit, we highlight that any mode-mixing performed by the initial beam splitter will necessarily reduce the \emph{local} single-mode squeezing, while at the same time generating some amount of two-mode squeezing. Thus, removing the beam splitter actually maximizes the squeezing on the phase shifters while, at the same time, it partially trivializes the problem to a single-mode one. \\

If we consider, instead, the other combination of parameters maximizing the $q$-dependent term, dubbed the \emph{optimal configuration} from now on, we obtain for $\mathcal{Q}_{22}$:
\begin{equation}
\label{eq:Q22opt}
       \mathcal{Q}_{22}^{opt} \ := \ \mathcal{Q}_{22} ( r, \theta = \frac{\pi}{2} , \phi = \frac{\pi}{4} , q=0 ) \  =   \ 2 \cosh^2 2 r  \cosh^2 2 x 
\end{equation}
The ratio between $\mathcal{Q}_{22}^{opt}$ over $\mathcal{Q}_{22}^{max}$ is:
\begin{equation}
    \frac{1}{4} \ \leq \ \dfrac{\mathcal{Q}_{22}^{opt}}{\mathcal{Q}_{22}^{max}} \ = \  \dfrac{\cosh^2 2 r \cosh^2 2 x }{\cosh^2 \left[ 2 ( r + x) \right]} \ = \ \frac{1}{4} \left( 1 + \dfrac{ \cosh [ 2 (r - x) ] } { \cosh [ 2 (r + x) ] } \right)^{2} \ \leq \ 1
\end{equation}
the lower bound being achieved in the limit of $r, x \to \infty$, while the upper bound is reached for either $r=0$ or $x=0$. Importantly, we see that the scaling of the QFI for the parameter $\lambda_{2}$ in the optimal configuration is the same as the scaling at its maximum, Eq.(\ref{eq:Q22max}) and it differs from it by at most a factor of $\frac{1}{4}$. Additionally, if in the best setting with $\phi = \theta = 0$ we do not assume to know $\lambda_{1}$, so that we can not impose $\gamma = 0$ to get the overall maximum configuration, in the worst-case scenario it could happen that $\gamma = \pi$ and we end up with:
\begin{equation}
\mathcal{Q}_{22}^{inf} \ \ = \ \ \mathcal{Q}_{22}( r, \theta = 0, \phi = 0, \gamma = \pi ) \ \ = \ \ 1 + \cosh [ 4 (r-x)]
\end{equation}
In this case, the ratio $\mathcal{Q}_{22}^{inf} / \mathcal{Q}_{22}^{opt}$ evaluates to $( \tanh 2x \tanh 2r - 1)^{2} \leq 1$. Thus the optimal configuration can even be more informative than the maximum configuration if we do not assume to know $\lambda_{1}$.

\section{Addressing sloppiness and quantum incompatibility}
\label{sec:III}

Let us now make a crucial observation: the optimal configuration also makes the estimation problem \emph{fully covariant}: indeed, both $\mathcal{Q}_{22}^{opt}$ in Eq.(\ref{eq:Q22opt}) and $\mathcal{Q}_{12}^{opt}$, whose value is:
\begin{equation}
    \mathcal{Q}_{12}^{opt} \ \ = \ \ 2 \cosh^{2} 2r + 2 \sinh^{2} 2r \sinh^{2} x 
\end{equation}
are now independent from both $\lambda_{1}$ and $\lambda_{2}$. \\

As a measure of the sloppiness in the various configurations, we can consider the determinant of the full matrix $\boldsymbol{\mathcal{Q}}$. Its value is exponentially growing with $r$ and $x$ both in the maximum and optimal configurations, but the ratio $\det \boldsymbol{\mathcal{Q}}^{max} / \det \boldsymbol{\mathcal{Q}}^{opt}$ is always bounded between $0$ and $2$ and it quickly goes to $0$ for large values of the initial squeezing $r$, whatever the value of $x$. This indicates that, as expected, the optimal configuration is better able to lift the sloppiness of the model independently of the energy introduced by the scrambler, quantified by $x$. \\

Let us now consider the quantum incompatibility of the parameters, computed through the Uhlmann curvature. 
Only off-diagonal entries can be non-zero and clearly $\mathcal{U}_{jk} = - \mathcal{U}_{kj}$. In our case, we get $\det \boldsymbol{\mathcal{U}} = \mathcal{U}^{2}_{12}$. The value of $\mathcal{U}_{12}$ can actually be written in a convenient closed form also for $q \neq 0$:
\begin{align}
    & \mathcal{U}_{12} ( \lambda_{2} ; r, \theta, \phi  , x , \gamma, q, \beta )  \notag
     \\ & = \ \ 2 \Big[ \cos ( \gamma + \theta) \cos 2 \phi \sin \theta  - 
     \cos\theta \sin(\gamma + \theta) \Big] \sinh 2 r \sinh 2 x  \ + 
     -   4 q^2 \cos^2 \phi \sin  2 (\beta - \lambda_{2} ) \sinh 2 x
 \end{align}
It is simple to check that $\mathcal{U}_{12}^{opt} = 0$, so that also the $\mathcal{R}$ parameter is zero in the optimal configuration and the parameters can be jointly estimated via the projection-valued measure diagonalizing simultaneously both the symmetric logarithmic derivatives. Moreover, the result is true \emph{for any value of $\lambda_{1}$ and $\lambda_{2}$} since, once again, the problem becomes covariant in the optimal configuration.

\section{Discussion and conclusions}
\label{sec:conclusions}
Before concluding our paper, it is worthwhile to consider the physical interpretation of the optimal configuration. Since we have $\phi = \frac{\pi}{4}$, the beam splitter is balanced. Additionally, for $\theta = \frac{\pi}{2}$, the input state after the beam splitter corresponds to a \emph{twin-beam state}, aside from a relative displacement and irrelevant local phases, which can be absorbed into an immaterial redefinition of $\lambda_1$. In this configuration, the local squeezing at the phase shifters is zero, while the entanglement between the modes is maximized.
This maximal entanglement might be connected to its effectiveness in decoupling the two parameters, $\lambda_1$ and $\lambda_2$. Entanglement plays a role in fully exploiting the two-mode space, allowing independent encoding of $\lambda_1$ and $\lambda_2$. In contrast, at maximal configuration, reducing the system to a single-mode problem might have constrained the independent degrees of freedom through which the two phase shifters can operate. However, further investigation with other models is necessary to draw robust and general conclusions on this matter.

In conclusion, we have conducted a detailed analysis of a sloppy continuous-variable quantum statistical model, specifically involving the encoding of two phase-shift parameters within the same arm of a Mach-Zehnder interferometer. Our results demonstrate that sloppiness can be corrected, effectively reducing quantum incompatibility to zero while preserving enhanced scaling of precision and maintaining the model's covariance with respect to the exact values of the parameters. Additionally, we have shown that concentrating quantum resources on the optical element to be estimated is generally the best strategy for enhancing precision. On the other hand, utilizing entanglement in the probe state proves more effective for estimating the two parameters independently.

Our results demonstrate that sloppy continuous-variable quantum statistical models can be effectively addressed, paving the way for quantum-enhanced metrology in biological samples and non-homogeneous crystals, which typically require a multi-parameter approach.

\begin{acknowledgments}
This work has been partially supported by Italian Ministry of Research 
and Next Generation EU via the PRIN-2022 project RISQUE (contract n. 2022T25TR3) 
and the PRIN-2022-PNRR project QWEST (contract n. P202222WBL).
\end{acknowledgments}

\begin{appendix}

\section{Continuous-variable systems and Gaussian states}
\label{appx:Gauss}
In continuous-variable quantum mechanics, one starts from the Fock space of $M$ modes and associates with each mode a pair of creation and annihilation operators satisfying the commutation rules $[ \hat{a}_j, \hat{a}_{k}^\dagger] = \delta_{jk}$ with $j,k \in \{ 1,...,M \}$. With our  conventions, the corresponding quadrature operators are given by $\hat{q}_j = \dfrac{\hat{a}^\dagger + \hat{a}}{\sqrt{2}}$, $\hat{p}_j = \dfrac{\hat{a}^\dagger - \hat{a}}{i\sqrt{2}}$, such that $[ \hat{q}_j , \hat{p}_k ] = i \delta_{jk}$. Those can be listed in an ordered vector of quadratures denoted by $\hat{\bm{R}} = \{ \hat{q}_{1}, \hat{p}_1 , ..., \hat{q}_M, \hat{p}_M \}$. Given a quantum state $\rrho$ of such a system, we define its first-moments vector and its covariance matrix (CM) as:
\begin{equation}
\label{eq:defCM}
   \vec{d}_{0} = \Tr[ \rrho \hat{\bm{R}}] \ , \ \ \left[{\sigma} \right]_{jk} \ = \ \frac{1}{2} \left\langle  \hat{R}_{j} \hat{R}_{k} + \hat{R}_{k} \hat{R}_{j} \right\rangle  - \left\langle \hat{R}_{j} \right\rangle \left\langle \hat{R}_{k} \right\rangle 
\end{equation}
where $\langle \hat{O} \rangle = \Tr [ \rrho \hat{O} ]$. Let us now introduce the multi-mode displacement operator $\hat{\bm{D}}(\vec{\Lambda})$, defined as:
\begin{equation}
 \hat{\mathbf{D}} \left( \vec{\Lambda} \right) \ \ := \ \ \exp{ \left[  -i \vec{\Lambda}^{T} \Omega \hat{\mathbf{R}} \right] }  
\end{equation}
where $\Omega := \bigoplus_{j=1}^{n} \boldsymbol{\omega}_{j}$ and $\boldsymbol{\omega}_{j}$ is a $2 \times 2$ matrix acting on the subspace $(\hat{q}_{j}, \hat{p}_{j})$ and equal to $i \sigma_{2}$, where $\sigma_{2}$ is the second Pauli matrix $\sigma_{2} = \left( \begin{array}{cc} 0 & -i \\ i & 0  \end{array} \right)$. Using this operator, we can represent each multi-mode state $\rrho$ by a complex function on phase-space, called the \emph{characteristic function} of the state:
\begin{equation}
    \label{eqdefcharfunct}
    \chi \left[ \Rh   \right] \left( \vec{\Lambda} \right) \ \ := \ \ \Tr \left[ \Rh \ \hat{\mathbf{D}} \left( \vec{\Lambda} \right)   \right]
\end{equation}
The Fourier transform of this function is again a function on phase space, but it is guaranteed to be real as a consequence of the self-adjointness of $\rrho$, and it is known as the \emph{Wigner function}. $\rrho$ is then called a \emph{Gaussian state} if its Wigner function is a Gaussian function; in that case, the covariance matrix and central first moments of the Gaussian will be exactly the CM $\sigma$ and the first-moments vector $\vec{d}_0$ of the state.

\section{Explicit expressions for the terms proportional to the initial displacement}
\label{appx:A}
The general formulas for the terms proportional to $q^2$ in the expressions for $\mathcal{Q}_{22}$ and $\mathcal{Q}_{12}$ are:
\begin{equation}
\begin{aligned}
     f_{22}(r, \beta, \theta,\phi, x, \gamma) \ =  &\  \sinh 2 r   \bigg[ \cos 2 \beta  \cos 2 \phi + \cos^2 \phi \Big( 2 \cosh ( 2 x )  \big\{ 2 \cos ( 2 \beta + \theta)  \sin \theta  \sin^2 \phi \  +  \\
        & + \ \sinh ( 2 x)  \big[ \cos \theta \cos(\gamma + \theta) + \cos ( 2 \phi) \sin \theta \sin (\gamma + \theta) \big] \big\} \ + \\
        & + \ \sinh ( 2 x ) \big\{ 4 \cos(\gamma + \theta) \sin \theta \sin^2 \phi \ + \\
        & + \ 2 \sinh ( 2 x) \cos ( \gamma - 2 \beta) \big[ \cos \gamma - 2 \sin\theta \sin(\gamma+\theta) \sin^2\phi \big]  \big\}  \Big)  \bigg] \ + \\
        & + \  \cosh 2r  \Big\{ 1+\cos^2\phi \ \big[ \cosh (4x )+\cos(\gamma-2\beta)\sinh (4 x  )  -1 \big] \Big\}
\end{aligned}
\end{equation}

\begin{equation}
\begin{aligned}
     f_{12}(r, \beta, \theta,\phi, x, \gamma) \ =  & \   \sinh 2 r  \bigg[ \cos 2 \beta \left( 2 \cosh^2 x \cos^2 \phi - 1 \right)  \ + \\
     & \ + \ 2 \sin \theta  \Big\{ 2 \left[ \cos(2\beta + \theta) \cosh^2 x - \sin (2 \beta + \theta) \sinh^2 x \right] \ + \\
    & \ - \ \sinh 2 x (\sin (\gamma + \theta) - 
            \cos (\gamma + \theta)) \Big\} \cos^2 \phi \sin^2 \phi \ + \\ 
    & \  + \ \sinh 2 x  \cos^2 \phi \cos \gamma \bigg] +  \cosh 2 r  \ + \\
    & + \ 2 \cosh 2 r  \sinh
        x  \cos^2 \phi \big[ \sinh x  + \cos (\gamma - 2 \beta) \cosh x \big] 
\end{aligned}
\end{equation}
The first expression in the maximum configuration ($\theta=\phi=\beta=\gamma=0$), which maximizes the $q$-independent term of $\mathcal{Q}_{22}$, becomes:
\begin{equation}
    f_{22}(r, x, \theta=\phi=\beta=\gamma=0) \ =  \  e^{2r +4x}
\end{equation}
Whereas in the optimal configuration it evaluates to:
\begin{equation}
\begin{aligned}
    & f_{22}(r, \beta, \theta = \frac{\pi}{2},\phi = \frac{\pi}{4}, x, \gamma) \ =  \   - \sinh ( 2 r )    \Big[  \cosh ( 2 x )  \sin ( 2 \beta ) +  \sinh (2 x  )  \sin \gamma   \Big]   \ + \\
        & + \  \cosh (2r)  \cosh (2x)\Big[ \cosh(2x)+\sinh(2x) \cos(\gamma-2\beta ) \Big]
\end{aligned}
\end{equation}
Setting $\beta = \gamma = 0$ in this last expression as well, we can compute the ratio between $f_{22}^{opt}$ and $f_{22}^{max}$:
\begin{equation}
    \dfrac{f_{22}^{opt}(\beta = \gamma = 0)}{f_{22}^{max}} \ = \ \dfrac{ e^{2x} \cosh (2r) \cosh(2x)}{e^{2r +4x}} \ = \ \dfrac{(1+e^{-4x})(1+e^{-4r})}{4}
\end{equation}
which is also lower bounded by $\frac{1}{4}$, as was the ratio $\mathcal{Q}_{22}^{opt}/\mathcal{Q}_{22}^{max}$ for $q=0$.

\end{appendix}

\bibliography{sloppyCVQSM.bib}

\end{document}